\documentclass[a4paper,conference]{IEEEtran}
\IEEEoverridecommandlockouts
\usepackage{cite}
\usepackage{amsmath,amssymb,amsfonts}
\usepackage{graphicx}
\usepackage{booktabs}
\usepackage{url}
\usepackage{pgfplots}
\pgfplotsset{compat=1.16}
\usepackage[hidelinks]{hyperref}
\usepackage{microtype}
\microtypesetup{factor=1100,stretch=25,shrink=30}

\IEEEpubid{\makebox[\columnwidth]{979-8-3195-3740-9/26/\$31.00~\copyright~2026 IEEE\hfill}
\hspace{\columnsep}\makebox[\columnwidth]{ }}

\begin{document}

\title{Evidence-Unit Fairness and the Limits of\\ Query-Adaptive Sparse-Dense Fusion in\\ Financial Document Retrieval}

\author{%
\IEEEauthorblockN{Chenyu Wu}
\IEEEauthorblockA{\textit{Duke University}\\
Durham, NC, USA \\
wuchenyu999@outlook.com}
\and
\IEEEauthorblockN{You Lin}
\IEEEauthorblockA{\textit{Independent Researcher}\\
New York, NY, USA \\
yl2356@cornell.edu}}

\maketitle

\begin{abstract}
Retrieval over financial filings is difficult because queries are short and acronym-heavy while the answer-bearing evidence sits inside long, table-dense documents. We study sparse-dense hybrid retrieval on FinDER, a benchmark of expert-annotated questions over corporate 10-K filings. Our first finding is methodological: if the retrieval unit is larger than the dense encoder's input window, the dense model never sees a large share of the labeled evidence, confounding comparison against a full-text sparse baseline. We measure this directly and reduce the imbalance with windows chosen for the encoder budget. On this corpus, fusing BM25 and a compact dense encoder improves reference-level Hit@10 by roughly 28 percent over either component, and training-free, untuned reciprocal rank fusion exceeds the equal-weight blend in an exploratory comparison. We then ask whether choosing the fusion weight per query helps: an oracle over the interpolation-weight grid shows headroom of 21.8 percent, yet none of the three lightweight adaptive routers (a score-confidence heuristic, a random forest over query features, and a ridge regressor over query embeddings) establishes a statistically reliable improvement over the fixed blend under company-grouped cross-validation with cluster-robust inference. Simple fusion is a strong baseline here, and we discuss why per-query weighting does not capture the available headroom.
\end{abstract}


\section{Introduction}
Question answering over corporate disclosures such as annual 10-K filings is a practical need in finance, and retrieval quality is the bottleneck. Financial machine learning already spans tasks such as distress prediction \cite{liu2023distress}, interpretable factor decomposition \cite{han2026factor}, firm-level investment analysis \cite{liu2020spillover}, and fraud and risk modeling in noisy, imbalanced transaction data \cite{xu2025discrimination,xu2026generative}. Our task is different: it asks a retrieval system to locate the filing evidence relevant to a financial question, returning the supporting sentence or table row and not only a score. A user asks a compact question, often using tickers and accounting shorthand, and the system must locate a specific passage or table inside a document running to hundreds of pages. Sparse lexical retrieval such as BM25 remains a strong baseline here because exact term and number matching matters, while dense retrieval adds semantic matching that helps when query and evidence share meaning but not words; combining the two, usually by fusing ranked lists, is now standard practice.

Two questions motivate this paper. First, are the two families compared on equal terms? A dense encoder truncates its input at a fixed token budget, so if the retrieval unit is a large table or multi-paragraph section, the encoder embeds only the opening fragment while a lexical index still matches the full text; on FinDER this is not marginal, and a large share of the labeled evidence begins beyond the point a compact encoder can read. Second, under a common windowing scheme, does adapting the fusion weight to each query improve retrieval, or is a single fixed weight enough?

We make the following contributions.
\begin{itemize}
\item We quantify an evidence-unit fairness problem on FinDER and reduce the truncation imbalance by indexing both retrievers over the same windows, with a reference-level mapping from each reference to its windows.
\item On the windowed corpus, both fusion methods clearly outperform either individual component, and a training-free rank fusion scores highest overall; the equal-weight blend's advantage over the individual components remains strong on the subset of labels whose references are literally contained in their windows.
\item We establish that real per-query headroom exists via an oracle weight analysis, then show that three lightweight adaptive routers do not convert that headroom into a reliable gain over the fixed blend under a company-grouped evaluation with cluster-robust testing.
\item We characterize automatic label quality, including where the mapping fails, and trace part of the difficulty to how table evidence is parsed rather than to retrieval.
\end{itemize}

\IEEEpubidadjcol
\section{Related Work}
BM25 and the probabilistic relevance framework it belongs to remain the reference point for lexical retrieval \cite{robertson1994,robertson2009}. Dense retrieval with dual encoders, trained with contrastive objectives, learns semantic matching and is competitive on many benchmarks \cite{karpukhin2020,reimers2019,wang2022e5}. Retrieval quality directly limits retrieval-augmented generation \cite{lewis2020}. Representation imbalance can amplify bias in generative recommendation \cite{fan2026crab}. Retrieval also serves as an auxiliary component outside text generation, including temporal retrieval for multi-modal popularity prediction \cite{lu2026m3tr}. Our empirical question is specific to financial-document retrieval.

A recurring practical concern is the retrieval unit: dense encoders have a bounded input length, and both retrieval effectiveness and the sparse-dense comparison depend on how documents are segmented into passages \cite{luan2021}. Late-interaction models such as ColBERT operate at token and passage granularity for this reason \cite{khattab2020}. Our fairness point is more specific than the general observation that encoders truncate: on FinDER the labeled evidence units, as first parsed, are much larger than the encoder budget and the evidence is often located deep inside a unit, so an uncorrected comparison measures truncation rather than semantics. We measure this effect, index both retrievers over common windows, and map references to those windows.

Hybrid systems combine lexical and dense signals. Reciprocal rank fusion combines ranked lists with a single rank offset and no per-query tuning, and is a robust default \cite{cormack2009}. A common alternative is a weighted score combination with a single interpolation weight. Selective component updates can reduce conflicts between forgetting and retention objectives in LLM recommendation \cite{chen2026cure}. Recent retrieval work, the closest prior method to ours, predicts the per-query weight from query representations with a classifier over discretized weight bins trained against each query's precomputed performance curve, with results broken down by language \cite{qahs2026}. We instead freeze the two retrieval components and train regressors, including one over query embeddings, on a continuous rank-based target within a single-domain financial-filings benchmark. Systems built for semi-structured collections plan the retrieval path before fusing and reranking what comes back \cite{tao2026grasp}. A parallel line asks not how to weight retrieval but whether to retrieve at all, judging that decision on utility, calibration, and cost together rather than accuracy alone \cite{qian2026active}. Our study shares that spirit but differs in emphasis: we first correct the evaluation confound, then test whether several lightweight adaptive strategies, including that embedding-based regressor, clear the fixed equal-weight blend under a company-grouped, leakage-controlled protocol.

FinDER is a benchmark for retrieval and retrieval-augmented generation over 10-K filings, with expert-annotated query and evidence pairs \cite{finder2025}; the original release evaluates single retrievers and a reranking stage, while we build a windowed corpus and a reference-level relevance set to study fusion and per-query weighting instead. A separate strand of work treats the filing itself as the object to improve, for instance using language models to make segment disclosures more complete and comparable across firms \cite{liu2026segment}; that work and ours share the constraint that what a system can find depends on how the disclosure was written and parsed. Related benchmarks control what a system may see \cite{zhu2026trading}, and larger memory alone need not improve behavior \cite{liu2026memory}. We apply that principle to retrieval units.

\section{Benchmark and Corpus Construction}
\subsection{FinDER}
FinDER provides expert-written queries paired with evidence drawn from 10-K filings of large United States companies, together with the filings themselves \cite{finder2025}. The queries are short and use financial shorthand, and the evidence spans both narrative passages and numeric tables. We use the released filings as the retrieval corpus and the released query and reference pairs as relevance labels.

\subsection{The evidence-unit fairness problem}
The filings are supplied as inline XBRL HTML, and a first parse into table and paragraph units yields large units: the median labeled evidence unit is about 5{,}000 characters, and the largest exceed one hundred thousand. Our dense encoder, described in Section IV, has a maximum input length of 512 tokens. That budget is not incidental: how much text a model reads is the main cost lever in a served pipeline and is traded against quality deliberately \cite{shen2026efficiency}, so a practitioner who picks a small encoder inherits its limit on purpose. For each labeled query and evidence unit we locate the earliest point at which any of the query's references begins. In 46.2 percent of these audited pairs that point lies at or beyond word 512, and in 55.2 percent beyond word 256, the tighter budget some compact encoders use. Both figures are conservative: they count whitespace-separated words rather than subword tokens, and they credit each pair with its earliest reference, so any remaining references in the same unit sit deeper still. A lexical index tokenizes the full unit, so BM25 can match evidence a truncating encoder never encodes; any comparison on these units is therefore biased toward the lexical side for a substantial fraction of references, and a weak dense result would say more about truncation than semantics.

\subsection{Windowing and relevance mapping}
We use windows of 160 words with 32 words of overlap as a conservative operating point under the encoder limit, not as an optimal retrieval unit. Across 497 filings, this yields 354{,}501 windows. Only 50 windows (0.014 percent) exceed 512 tokens, and none is a mapped labeled-evidence window.

We then map each labeled reference to its windows automatically: for a reference we locate its text inside the parsed unit FinDER associates with the query, using a prefix-key search over progressively shorter keys with a similarity check to disambiguate repeated occurrences, then mark every overlapping window as relevant, so a reference spanning several windows marks all of them. References that cannot be located are excluded. This procedure resolves 5{,}386 of 6{,}121 references, covering 5{,}082 of 5{,}703 queries (89.1 percent). We keep provenance from each query to each reference to each window, which lets us evaluate at the level of references rather than raw windows.

\subsection{Label quality}
Because the mapping is automatic and approximate, we report deterministic diagnostics rather than a verified precision figure. A literal-containment diagnostic succeeds for 30.3 percent of references, meaning the normalized reference text is a literal substring of its mapped windows; 1{,}518 queries have all references literally contained (the literal-containment subset). A looser text-overlap score places 81.8 percent of references above 0.5 and 44.7 percent above 0.7. Neither diagnostic is a precision bound: a correct mapping can fail literal containment, and high overlap does not guarantee correctness.

Qualitative inspection suggests the mapping failures concentrate in financial-statement tables, where the inline-XBRL parse can separate a heading from its numeric rows. For example, a query asking for a firm's current-year SG\&A-to-net-sales ratio maps to a window holding only the statement title and the standard note that the accompanying notes are integral, with none of the numeric rows; no retriever can recover the answer from such a window, so the failure lies in how the evidence unit was constructed, not in ranking. We return to its consequences in the discussion.

\section{Retrieval Methods and Evaluation Protocol}
\subsection{Retrievers and fusion}
The lexical retriever is BM25, implemented with the \texttt{bm25s} library using its Lucene scoring variant, $k_1=1.5$ and $b=0.75$, tokenizing on alphanumeric terms without stopword removal, indexed over all windows. The dense retriever is a compact retrieval-trained encoder, e5-small-v2 \cite{wang2022e5}, applied with its query and passage prefixes and a maximum sequence length of 512 tokens, with exact inner-product search over L2-normalized embeddings \cite{johnson2019}; each retriever returns its top 100 windows per query. We consider two fusion strategies. Reciprocal rank fusion combines the two ranked lists with a rank offset $k=60$ and requires no per-query tuning \cite{cormack2009}. Weighted fusion combines per-query min-max normalized scores with an interpolation weight $\alpha$, where $\alpha=1$ is pure BM25 and $\alpha=0$ is pure dense. Both strategies score the union of the two top-100 candidate lists. In weighted fusion a candidate absent from one list receives a normalized score of zero for that component; in RRF an absent candidate simply contributes no reciprocal-rank term from that list.

\subsection{Oracle and adaptive routers}
To measure how much a per-query weight could help, we compute a grid oracle that selects, for each query, the weight on the evaluated grid $\{0.0,0.05,\dots,1.0\}$ that maximizes that query's score. The oracle is not a deployable method; it is an empirical upper bound for any router restricted to this same grid, not over every conceivable per-query linear weighting.

We evaluate three practical routers that predict a per-query weight. The first is a model-free heuristic: each retriever's top-one score is separately min-max scaled with parameters fit on the training split, and the weight is $\alpha = b/(b+d)$ from the scaled scores $b$ and $d$, defaulting to 0.5 when both are zero. Confidence-based escalation of this kind is a familiar pattern in moderation systems \cite{xin2026driftguard}. The second is a random forest \cite{breiman2001} regressor over twelve inexpensive query-surface features\footnote{Character length, word length, digit count, uppercase-token count, ticker-mention count, parenthesized-ticker flag, financial-acronym count, comparison-word count, financial-statement-word count, dollar-sign flag, percent-sign flag, and average word length.}; related prompt indicators have been examined in generative-model settings \cite{ainiwaer2026indicator}, although our task and target differ. The third is a ridge regressor over the query embedding produced by the dense encoder, with predicted weights clipped to $[0,1]$ before fusion. The two learned routers are trained to predict a rank-based target: for each query we take the best rank achieved by any relevant window under each retriever, with rank 101 if none of its relevant windows appears in the top 100, and compute $r_d/(r_b+r_d)$ from the dense rank $r_d$ and BM25 rank $r_b$, excluding queries where neither retriever finds a relevant window in its top 100. The random-forest grid is $200$ trees with \texttt{max\_depth} in $\{3,6,\texttt{None}\}$ and \texttt{min\_samples\_leaf} in $\{5,10,20\}$; ridge regularization $\alpha$ is selected from $\{1,10,50,100,300,1000\}$, both by validation Hit@10, using scikit-learn \cite{pedregosa2011}.

\subsection{Metrics and protocol}
Our primary metric is reference-level Hit@10: a reference counts as hit when any of its windows appears in the top ten, and a query score averages over its references. We also report window-level Recall@10 as a secondary metric, since it can reward retrieving several overlapping windows of a single long reference.

We evaluate with grouped cross-validation: queries are grouped by the company of their evidence, and folds are drawn so no company appears in both training and test, which prevents a router from seeing a filing's surface features at training time and being tested on the same filing. We use five folds and twenty shuffled repetitions of the same 489 companies, confirmed at runtime to differ across repetitions; since repetitions are not independent replications, we treat the company as the unit of independence throughout. We report two fixed-fusion baselines that must be kept distinct: the equal-weight blend uses $\alpha=0.5$ with no tuning, while the tuned blend selects $\alpha$ on a validation split drawn from the training companies, as do the router hyperparameters.

For the router comparisons, each query's out-of-fold predictions were averaged across the twenty repetitions before inference, so every query contributes once. Significance uses a company-clustered bootstrap with 10{,}000 replicates over the 489 company clusters, the unit of independence under grouped cross-validation. For a comparison declared in advance in a single direction we use a fixed tail regardless of the observed sign, and a two-sided test for exploratory comparisons; a company-level sign-flip permutation test served as a cross-check for the primary router comparisons and agreed with the bootstrap. A router is judged to help only if it beats the equal-weight blend on reference-level Hit@10 with a one-sided clustered $p$ below a Bonferroni-adjusted threshold of $0.05/3 \approx 0.0167$, correcting for the three adaptive methods evaluated over the project (heuristic, random forest, ridge), each fixed before that router's own final evaluation and tested in sequence rather than planned jointly at the outset. None of the routers clears even the unadjusted 0.05 level, so the conclusion does not depend on the exact adjustment; in a different setting, deterministic oracles expose agents learning from failures that never happened \cite{wang2026phantom}.

\section{Results}
\subsection{Fusion outperforms either retriever}
Table~\ref{tab:main} reports reference-level Hit@10 and window-level Recall@10. Neither component is strong alone, and the two are close to each other under the common windowing design, with BM25 at 0.1388 and the dense model at 0.1415. Fusion is a large gain over either: the equal-weight linear blend reaches 0.1802, an improvement of 29.9 percent over BM25 and 27.4 percent over the dense model. Reciprocal rank fusion, combining the two ranked lists without score calibration, per-query weighting, or training, scores best at 0.1892, exceeding the equal-weight blend by 0.0090 in an exploratory two-sided comparison under the same company-clustered bootstrap (95 percent interval $[0.0030, 0.0152]$, $p=0.0046$), so the highest-scoring configuration here is also the simplest. We treat this as exploratory because rank fusion was not part of the advance router-decision rule.

\begin{figure}[t]
\centering
\begin{tikzpicture}
\begin{axis}[
    width=\columnwidth,
    height=3.6cm,
    xlabel={Interpolation weight $\alpha$ (1 = BM25, 0 = dense)},
    ylabel={Hit@10 (ref)},
    xmin=0, xmax=1, ymin=0.135, ymax=0.19,
    grid=both, grid style={gray!20},
    tick label style={font=\footnotesize},
    label style={font=\footnotesize},
]
\addplot[thick, mark=*, mark size=1pt] coordinates {
(0.00,0.1415)(0.05,0.1450)(0.10,0.1495)(0.15,0.1524)(0.20,0.1561)
(0.25,0.1603)(0.30,0.1635)(0.35,0.1701)(0.40,0.1764)(0.45,0.1797)
(0.50,0.1802)(0.55,0.1781)(0.60,0.1732)(0.65,0.1668)(0.70,0.1622)
(0.75,0.1555)(0.80,0.1530)(0.85,0.1480)(0.90,0.1442)(0.95,0.1412)(1.00,0.1386)
};
\end{axis}
\end{tikzpicture}
\caption{Reference-level Hit@10 as a function of the fixed interpolation weight. The optimum is at equal weight and the region around it is flat. The $\alpha=1$ endpoint (0.1386) differs from the BM25-only row in Table~\ref{tab:main} (0.1388) by about one query out of 5{,}082, because this sweep re-derives the top ten from the normalized union of both retrievers' candidate sets rather than reading BM25's own precomputed ranking directly, and BM25's integer-valued scores create ties the two procedures can break differently.}
\label{fig:sweep}
\end{figure}
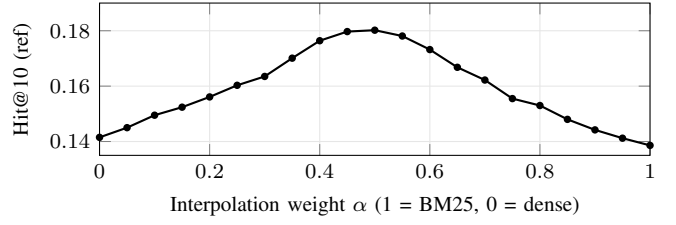

The linear blend is not sharply tuned. Fig.~\ref{fig:sweep} shows reference-level Hit@10 as a function of the interpolation weight. The curve is a broad inverted U with a peak at $\alpha=0.5$ and a flat region between 0.45 and 0.55, so an equal weight is both simple and near optimal.

\begin{table}[t]
\caption{Retrieval quality on the windowed FinDER corpus. The primary metric is reference-level Hit@10; window-level Recall@10 is secondary. Higher is better. Best-performing deployable method in bold; the oracle is an empirical upper bound over the evaluated interpolation grid, not a deployable method.}
\label{tab:main}
\centering
\footnotesize
\setlength{\tabcolsep}{4pt}
\renewcommand{\arraystretch}{0.98}
\begin{tabular}{@{}lcc@{}}
\toprule
Method & Hit@10 (ref) & Recall@10 (win) \\
\midrule
BM25 only & 0.1388 & 0.0596 \\
Dense only (e5-small-v2) & 0.1415 & 0.0584 \\
Fixed fusion, tuned $\alpha$ & 0.1785 & 0.0764 \\
Fixed fusion, $\alpha=0.5$ & 0.1802 & 0.0770 \\
\textbf{Reciprocal rank fusion} & \textbf{0.1892} & \textbf{0.0833} \\
\midrule
Heuristic router & 0.1641 & 0.0699 \\
Random forest router & 0.1801 & 0.0773 \\
Ridge router (embeddings) & 0.1824 & n/a \\
\midrule
Grid oracle, best $\alpha$ per query & 0.2195 & n/a \\
\bottomrule
\end{tabular}
\end{table}

\subsection{Headroom exists but routers do not capture it}
A grid oracle that picks the best linear weight per query over the evaluated grid reaches 0.2195, 21.8 percent above the equal-weight blend. Room to improve therefore exists in principle for per-query weighting, but two considerations temper that conclusion. First, 77.6 percent of queries are not answered in the top ten under any weight, so the achievable gain is concentrated on the minority of queries any weighting can reach. Second, the training-free rank fusion already closes part of this gap without any per-query adaptation, raising the bar a router would have to clear to be useful.

The three routers do not realize the headroom. Table~\ref{tab:sig} reports exploratory two-sided company-clustered comparisons against the equal-weight linear blend, the baseline fixed in advance; the decision rule itself uses the one-sided, prespecified-direction $p$ against the 0.0167 threshold from Section IV. The random-forest router shows no evidence of improvement over the blend, with a difference of $-0.00004$ whose interval spans zero, favoring the blend in thirteen of twenty repetitions. The heuristic router is significantly worse, by 0.0161. The ridge router over embeddings gives a small positive estimate of $+0.00220$, but its interval includes zero and its one-sided $p$ of 0.0555 does not clear 0.0167. By the rule fixed in advance, none of the routers is judged to help; only the untuned rank fusion differs from the blend reliably, and in the positive direction.

\begin{table}[t]
\caption{Comparisons against the equal-weight ($\alpha=0.5$) blend at reference-level Hit@10, company-clustered inference with two-sided $p$. Positive favors the alternative. Only reciprocal rank fusion differs reliably in the positive direction.}
\label{tab:sig}
\centering
\footnotesize
\setlength{\tabcolsep}{4pt}
\renewcommand{\arraystretch}{0.98}
\begin{tabular}{@{}lccc@{}}
\toprule
Alternative $-$ blend & Difference & 95\% CI & $p$ \\
\midrule
Reciprocal rank fusion & $+0.00905$ & $[+0.0030, +0.0152]$ & 0.005 \\
Ridge router & $+0.00220$ & $[-0.0005, +0.0050]$ & 0.111 \\
Random forest router & $-0.00004$ & $[-0.0022, +0.0019]$ & 0.970 \\
Heuristic router & $-0.01613$ & $[-0.0220, -0.0103]$ & $<0.001$ \\
\bottomrule
\end{tabular}
\end{table}

\subsection{Where the retrievers disagree}
The two components have complementary strengths that vary by query category. Table~\ref{tab:cat} breaks reference-level Hit@10 down by the eight FinDER query categories. BM25 leads on accounting, financials, and legal queries, which turn on exact terms, defined language, and specific figures, while the dense encoder leads on company-overview, risk, and shareholder-return queries, which are more often phrased descriptively. The gaps are not small: dense retrieval is 29 percent better than BM25 on risk queries and 65 percent better on shareholder-return queries, while BM25 is more than twice as accurate on financials, the kind of regime-dependent split that aggregate benchmarks can conceal \cite{wang2026regime}.

This is the pattern a per-query router might try to exploit, yet the equal-weight blend already absorbs most of it. Fusion matches or beats both components in seven of the eight categories, and in the eighth, shareholder return, it trails the better single retriever by less than half a point. A category-based hard-selection baseline, which always returns whichever single retriever is best on average for a query's category, reaches only 0.158, below the 0.180 of the untuned equal-weight blend. This tests only a hard switch between the two pure retrievers, not a continuous, category-conditioned fusion weight, so it does not rule out every possible category-aware method. It does show that the most direct way of using category information is already dominated by the fixed blend, consistent with, though not a full explanation of, the failure of the lightweight routers we test.

\begin{table}[t]
\caption{Reference-level Hit@10 by FinDER query category, with the number of evaluated queries $n$. The better single component per row is in bold. The equal-weight blend matches or beats both components in every category except shareholder return.}
\label{tab:cat}
\centering
\footnotesize
\setlength{\tabcolsep}{4pt}
\renewcommand{\arraystretch}{0.98}
\begin{tabular}{@{}lcccc@{}}
\toprule
Category & $n$ & BM25 & Dense & Blend $\alpha{=}0.5$ \\
\midrule
Accounting         & 482  & \textbf{0.207} & 0.150 & 0.227 \\
Company overview   & 1029 & 0.155 & \textbf{0.187} & 0.222 \\
Financials         & 756  & \textbf{0.079} & 0.038 & 0.082 \\
Footnotes          & 839  & \textbf{0.112} & 0.104 & 0.141 \\
Governance         & 626  & \textbf{0.095} & 0.087 & 0.113 \\
Legal              & 466  & \textbf{0.182} & 0.157 & 0.210 \\
Risk               & 456  & 0.197 & \textbf{0.254} & 0.296 \\
Shareholder return & 428  & 0.134 & \textbf{0.222} & 0.217 \\
\bottomrule
\end{tabular}
\end{table}

\subsection{Sensitivity to label quality}
Because the labels are automatic, we repeat the key comparisons on the 1{,}518 queries whose references are literally contained in their windows, the subset least likely to be mislabeled. The central result strengthens: the equal-weight blend reaches 0.178 here, against 0.149 for the dense model and 0.127 for BM25, gains of 19.5 percent and 39.9 percent. Rank fusion retains a similar point estimate, at $+0.0092$ over the blend, but the smaller-subset interval includes zero (two-sided $p=0.11$), so we do not claim a reliable rank-fusion advantage on this subset. The router picture is unchanged in substance: the random-forest router remains at $-0.00125$ against the blend, and the ridge router shrinks to $+0.00026$, effectively zero. Label-quality sensitivity for the ridge router is therefore mixed, positive on the looser subset and near zero on the literal subset, with no subset establishing a reliable advantage.

A secondary, exploratory observation: tuning the fixed weight on a small validation split can slightly underperform simply setting it to 0.5 (difference $-0.00168$); we report this as suggestive, not confirmed, since it is an uncorrected two-sided comparison.

\section{Discussion}
In this FinDER and e5-small-v2 setting, the results support a simple recommendation: after the components are compared on equal terms, fusion is a strong baseline, and rank fusion obtains the highest score of any method in our study. None of the three lightweight per-query routers establishes a statistically reliable improvement over the equal-weight blend; this does not mean adaptive routing cannot help in general, only that the evaluated routers did not convert the surface features and compact embedding signal we tested into a reliable improvement over the fixed blend here, even though the grid oracle shows that per-query headroom exists. We did not pursue a per-query attribution of the random-forest predictions, since the router never separated from the blend in the first place, though local importance methods for tree ensembles \cite{liang2025localmdi} are the right instrument once there is a signal worth explaining.

Simplicity also has an operational value the tables do not show. Retrievers often sit inside tool-integrated pipelines where downstream components may not control retrieved inputs \cite{jiang2026agentic} or observe upstream objectives and provenance \cite{li2026dangerous}. A fusion rule with no learned component is easier to audit than a router whose behavior depends on a fitted model and the split it was fitted from.

Two properties of the setting explain much of the difficulty. First, a large fraction of queries are unreachable in the top ten under any weight, so the effective sample a router could help is small, and small effects are hard to establish under a leakage-controlled protocol. Second, part of the unreachability is a labeling and parsing artifact, not a retrieval failure: the heading-only mapping of financial-statement tables means some numeric queries' answer-bearing rows may not be present in any retrievable window, so no method could reach them. That is a measurement limitation, not a modeling one, and a related pattern shows up elsewhere, where a recorded field ends up encoding how the data was produced rather than what it was meant to capture \cite{liang2025adherence}. Better table-aware parsing that keeps a heading with its rows is therefore a promising direction for raising the retrieval ceiling.

The same argument applies above any one application: automated research pipelines increasingly assemble separate retrieval, reading, and synthesis stages \cite{kong2026autoresearch}, and a first stage that silently drops evidence passes that loss to everything built on top of it, where it is harder to see.

\subsection{Practical guidance}
In this FinDER and e5-small-v2 setting, our results point to a short set of choices for retrieval over financial filings. Reciprocal rank fusion is a sensible starting point: it scored highest among the methods we tested, needs no training, and its usual offset worked well, so there is little to tune. Do not assume lightweight per-query routing will improve on a fixed blend: a confidence heuristic, a forest over query features, a regressor over embeddings, and a hard switch by query category were all tested, and none of the tested adaptive methods establishes a reliable improvement over the fixed blend. Put that effort into how the evidence is cut instead: matching the retrieval window to the encoder budget reduced a substantial source of evaluation imbalance here, and keeping a financial-statement heading with its numeric rows could make this class of table answers retrievable. Last, measure at the level of the labeled reference and split the data by company, since scoring loose windows or letting one firm sit on both sides of a split makes a weak system look stronger than it is.

\subsection{Limitations}
The relevance labels are produced automatically, so we report deterministic diagnostics rather than verified precision. We used one compact dense encoder for reproducibility, and the numeric levels are specific to it, reflecting tight compute and privacy budgets \cite{venkatesh2026dpprox} and work on small model efficiency \cite{liu2026howllms}. We did not compare segmentation schemes. Other window sizes, paragraph boundaries, or table structure could affect context, duplication, candidate competition, and absolute scores. Windowing enlarges the candidate pool and gives references spanning several windows more chances to enter the top ten. Results cover 5{,}082 queries with resolved references. The 621 unresolved queries may differ systematically. The learned routers use one rank-based target, and a target aligned directly with Hit@10 could produce different results. Our routers are lightweight, and heavier models could behave differently.

\section{Conclusion}
We studied sparse and dense retrieval on financial filings and found that controlling retrieval unit length matters on FinDER. With windows of 160 words, simple fusion is a strong baseline, and training-free rank fusion scores best in our study. Real per-query headroom exists over the linear blend, yet none of the three tested routers establishes a statistically reliable improvement over it under a company-grouped, cluster-robust evaluation. For practitioners, untuned rank fusion is a sensible default here, and our results suggest the next gains may come from better evidence-unit construction rather than lightweight per-query weighting.

\end{document}